\documentclass[twocolumn,showpacs,showkeys,pra]{revtex4}
\usepackage{amsmath}
\usepackage{longtable}
\usepackage{graphicx}
\usepackage{dcolumn}
\usepackage{bm}

\newcommand{\be}{\begin{equation}}
\newcommand{\ee}{\end{equation}}
\newcommand{\ben}{\begin{eqnarray}}
\newcommand{\een}{\end{eqnarray}}

\begin{document}
\draft

\title{Entanglement and the Quantum Brachistochrone Problem}
\author{A. Borras$^1$, C. Zander$^2$,
A.R. Plastino$^{1,2,3}$\footnote{Corresponding Author:
arplastino@maple.up.ac.za}, M. Casas$^{1}$, and A. Plastino$^{3}$}

\address{$^1$Departament de F\'{\i}sica and IFISC-CSIC,
Universitat de les Illes Balears, 07122 Palma de Mallorca,
Spain \\ $^2$Physics Department, University of Pretoria,
Pretoria 0002, South Africa \\
$^3$National University La Plata-CONICET, C.C. 727, 1900 La Plata,
Argentina }

\date{\today}

\begin{abstract}
Entanglement is closely related to some fundamental features
of the dynamics of composite quantum systems: quantum entanglement
enhances the ``speed" of evolution of certain quantum states,
as measured by the time required to reach an orthogonal state.
The concept of ``speed" of quantum evolution constitutes an
important ingredient in any attempt to determine the fundamental
limits that basic physical laws impose on how fast a physical system can
process or transmit information. Here we explore the relationship
between entanglement and the speed of quantum evolution in the context
of the quantum brachistochrone problem. Given an initial and a final
state of a composite system we consider the amount of entanglement
associated with the brachistochrone evolution between those states,
showing that entanglement is an essential resource to achieve the
alluded time-optimal quantum evolution.


\end{abstract}



\maketitle



 The celebrated brachistochrone problem has played a
distinguished role in the history of mechanics \cite{M89}. A quantum
dynamical optimization problem akin to the classical brachistochrone
one has attracted the attention of several researchers in recent
years \cite{BH06,CHKO06,BDJM07,CHKO07}. This problem deals with the
Hamiltonian generating the optimal quantum evolution
$|\psi(t)\rangle$ (in the sense of the one requiring the shortest
time $\tau$) between two prescribed states $|\psi_I\rangle$ and
$|\psi_F\rangle$. It has been recently established that the
entanglement features of quantum states are important in connection
with dynamical optimization problems
\cite{GLM03a,GLM03b,BCPP05,BCPP06}. Previous research done on this
subject has been focused upon the study of the time needed to reach
an orthogonal state by systems evolving under a given Hamiltonian
\cite{ZPPC07}. The details of these kind of studies depend strongly
on the particular Hamiltonian considered, and each case requires
a separate treatment (see, for instance, \cite{BCPP05,BCPP06}).
The aim of the present contribution is to
revisit the connection between entanglement and optimal evolution
from the different (but related) point of view provided by the quantum
brachistochrone problem. An important advantage of the brachistochrone
approach is that it allows for a more general and unified
investigation of the alluded connection, going beyond the
separate analysis of individual cases.

\section{Role of entanglement in time-optimal quantum evolution}\label{chapt4sec6}

 As already mentioned, our aim is to explore the
 connection between entanglement and the quantum brachistochrone
 evolution of a bipartite system constituted by two subsystems
 $A$ and $B$ (with associated Hilbert spaces of dimensions
$N_A$ and $N_B$, respectively). We want to quantify the
 amount of entanglement involved when implementing the evolution of
 an initial state $|\psi_I\rangle$ to a final state $|\psi_F\rangle$ in
 the shortest possible time. Following \cite{BH06}, we are going to
 consider the optimal evolution under the constraint
that the difference between the maximum and minimum eigenenergies of
the Hamiltonian generating the unitary transformation
$|\psi_I\rangle \rightarrow |\psi_F\rangle =
e^{\frac{iH\tau}{\hbar}}|\psi_I\rangle$ be less or equal to a given
constant energy $2\omega$. This constraint is imposed due to the
following reason: if the differences between the eigenenergies of
the Hamiltonian are arbitrarily large, then it is easy to implement
a quantum evolution connecting the alluded states and taking a time
$\tau$ that can be made arbitrarily small. The optimal time
evolution is given by  \cite{BH06}

\begin{eqnarray}
|\psi(t)\rangle &=& \left[ \cos \left( \frac{\omega t}{\hbar}\right) -
\frac{\cos\frac{1}{2}\theta}{\sin\frac{1}{2}\theta}\sin\left(
\frac{\omega t}{\hbar}\right)\right]
|\psi_I\rangle\cr && + \frac{1}{\sin\frac{1}{2}\theta}\sin\left(
\frac{\omega t}{\hbar}\right)|\psi_F\rangle.
\end{eqnarray}

\noindent This expression satisfies $|\psi(0)\rangle =
|\psi_I\rangle$ and $|\psi(\tau)\rangle = |\psi_F\rangle$,
where

\begin{equation}
\tau = \frac{\hbar \theta}{2\omega}.
\end{equation}

\noindent To determine the parameter $\theta$, the final state
$|\psi_F\rangle$ has to be written in the form

\begin{equation}
|\psi_F\rangle = \cos\frac{1}{2}\theta|\psi_I\rangle+
e^{i(\phi+\pi/2)}\sin\frac{1}{2}\theta|\overline{\psi}_I\rangle,
\end{equation}

\noindent where $|\overline{\psi}_I\rangle$ is a state orthogonal to
the initial state $|\psi_I\rangle$ contained in the two-dimensional
subspace (of the full Hilbert space) spanned by the initial and
final states. Since both $|\psi_I\rangle$ and $|\psi_F\rangle$ are
specified, the values of $\phi$ and $\theta$ can be regarded as
known, the latter being the angle of separation of these two states.
We are going to focus upon the case where the initial and final
states are orthogonal, thus having $\theta = \pi$ and,

\begin{eqnarray}
\tau &=& \frac{\pi \hbar}{2 \omega} \cr |\psi(t)\rangle &=& \cos
\left( \frac{\omega t}{\hbar}\right) |\psi_I\rangle + \sin\left(
\frac{\omega t}{\hbar}\right)|\psi_F\rangle.
\end{eqnarray}

\noindent From the point of view of the physics of information and
computation this case is the most interesting because the evolution
of a computer devise into a state orthogonal to the initial one can
be identified with an elementary information processing step
\cite{ML98}.

 In order to assess how much entanglement is involved in the
brachistochrone evolution we are going to compute the time-average
of the entanglement $\mathcal{E}[\psi(t)]$ during the optimal time
evolution from $|\psi_I\rangle \rightarrow |\psi_F\rangle$,

\begin{equation}
\langle \mathcal{E} \rangle \,\,=\,\,
\frac{1}{\tau}\int_0^{\tau}\mathcal{E}(t)\mathrm{d}t.
\end{equation}

\noindent The idea of time averaged entanglement has
been recently discussed by several researchers and
was found to constitute a useful concept to study a
variety of problems (see, for instance, \cite{VN07}
and references therein). To facilitate the computations
we use the linear entropy as entanglement measure,

\be \mathcal{E}[\psi(t)] \, = \, \frac{N_A}{N_A-1}
\Bigl[ 1 - Tr \left(\rho_A^2
\right) \Big], \ee

\noindent where $\rho_A$ is the marginal density matrix associated
with one of the two subsystems (the one with the Hilbert space
of lower dimension), $\rho_A \, = \, Tr_B[\rho(t)]$,
and the global density matrix describing the bi-partite
system is $ \rho(t) = |\psi(t)\rangle \langle\psi(t)|$.
Since the entanglement $\mathcal{E}(t)$ depends upon time
only through the quantity $\xi = \frac{\omega t}{\hbar}$, the
integral giving the time-averaged entanglement can be put under the
guise

\be \langle \mathcal{E} \rangle \, = \,
\frac{2}{\pi} \int_0^{\pi/2}
\mathcal{E}(\xi) \, d\xi,\ee

\noindent implying that $\langle \mathcal{E} \rangle$
is independent of $\omega$ and since
$\tau = \frac{\pi \hbar}{2\omega}$, this means that
$\langle \mathcal{E} \rangle$ is also independent
of the absolute time taken.

\noindent Highly asymmetric states can evolve optimally without
entanglement. This is clearly illustrated by the extreme case
corresponding to factorizable initial and final states of the
form

\ben \label{defacto} |\tilde \psi_I\rangle \, &=& \, |\phi_1 \rangle
\otimes |\phi_0 \rangle \cr |\tilde \psi_F\rangle \, &=& \, |\phi_2
\rangle \otimes |\phi_0 \rangle, \een

\noindent where one of the subsystems (in this case,
the second subsystem) is in the same state $|\phi_0
\rangle$ at the beginning and at the end of the process. It is plain
that in such circumstances the brachistochrone evolution can be
implemented without entanglement. Indeed, the optimal evolution is
given by the time dependent, separable state

\be \label{going_solo}
|\tilde \psi(t)\rangle \, = \, |\phi(t) \rangle \otimes |\phi_0
\rangle, \ee

\noindent $|\phi(t) \rangle $ being the optimal evolution of the
first subsystem connecting the states $|\phi_1 \rangle $ and
$|\phi_2 \rangle$. It is also clear that in this highly asymmetric
setting only one of the subsystems (the first one)
is evolving and, consequently, we
are essentially dealing with the evolution of a single system. That
is, the composite nature of the total system plays no role
in an evolution like (\ref{going_solo}). This is
fully consistent with similar results reported by
Giovannetti, Lloyd and Maccone in \cite{GLM03a}.
There it was pointed out that, for composite a system
with non-interacting subsystems, assymetric non-entangled
states in which all the energetic resources are concentrated
in one subsystem can saturate the fundamental lower bound
for the time required to reach a state orthogonal
to the original one. On the contrary, symmetric non-entangled
states with evenly shared energetic resources do not
saturate the alluded bound.

{\it States of the form (\ref{defacto}) are the only
pairs of orthogonal initial and final states such that the time
averaged entanglement of the concomitant optimal evolution vanishes}.
This can be seen as follows. Suppose that the time averaged
entanglement is zero. That implies that the entanglement is zero at
all times $t\in[0,\tau]$. In particular, the initial and the final
states must be separable. Since they are also orthogonal we can
assume, without loss of generality, that the initial and final
states are, respectively, of the form
$|\psi_I\rangle=|0\rangle\otimes|\phi_r\rangle$ and
$|\psi_F\rangle=|1\rangle\otimes|\phi_s\rangle$. Now, the time
dependent state must be separable also at all intermediate times
$t\in[0,\tau]$. In particular, it must be separable at the time
corresponding to $\frac{\omega t}{\hbar}=\frac{\pi}{4}$. That is,
the state $(1/\sqrt{2})(|0\rangle\otimes|\phi_r\rangle +
|1\rangle\otimes|\phi_s\rangle)$ has to be separable. But for this
state we have $\mathcal{E}=1-|\langle  \phi_r| \phi_s \rangle|^2$.
Consequently, the entanglement of this state will be zero only if
$|\langle  \phi_r| \phi_s \rangle|=1$, and so the pair of initial
and final states is of the form (\ref{defacto}).

\section{Two-Qubit Systems}
 The features of the optimal evolution which are associated with
 the composite nature of the system (such as the role of
 entanglement) are more clearly seen when we study the evolution of
 symmetric states. For this reason we are now going to pay special
 attention on the optimal evolution of two-qubit systems associated
 with symmetric initial and final states.

 First we are are going to consider three specific cases of optimal
 evolution of symmetric sates of two qubit systems associated,
 respectively, with the following three pairs of initial and final
 states,

\begin{itemize}
\item (i) $\,\,\,\,\,$ $|00\rangle \rightarrow
\frac{\cos\alpha}{\sqrt{2}}\{|01\rangle+|10\rangle\}+
\sin\alpha |11\rangle$
\item (ii) $\,\,\,\,\,$ $\frac{1}{\sqrt{2}}\{|01\rangle+
|10\rangle\} \rightarrow
\frac{1}{\sqrt{2}}\{|00\rangle+|11\rangle\}$
\item (iii) $\,\,\,\,\,$ $\frac{1}{\sqrt{2}}\{|00\rangle -
i|11\rangle\} \rightarrow
\frac{1}{\sqrt{2}}\{i|00\rangle-|11\rangle\}$.
\end{itemize}

\noindent Let us discuss each example separately:

\noindent (i). In this case we have the evolution of a separable
state into one of a family of possible final states parameterized by
the real parameter $\alpha$. According to the value of this
parameter we may have for the final state either another separable
state ($\alpha=\frac{\pi}{2}$), an intermediately entangled state
($\alpha\in (0,\frac{\pi}{2})$) or a maximally entangled state
($\alpha=0$). Let $\xi = \frac{\omega t}{\hbar}$. Then the time
dependent, optimally evolving state, the concomitant entanglement,
and its time average are respectively given by,

\begin{eqnarray}
|\Psi(\xi,\alpha) \rangle &=& \cos{\xi} \, |00\rangle + \sin{\xi} \,
\frac{\cos{\alpha}}{\sqrt{2}} \, [|01\rangle + |10\rangle ] \cr
&+&  \sin{\xi} \, \sin{\alpha} \, |11\rangle, \cr
\mathcal{E}(\xi,\alpha)  &=& {\left( {\cos (\alpha )}^2\,{\sin (\xi
)}^2 -
\sin (\alpha )\,\sin (2\,\xi ) \right) }^2, \cr
\langle \mathcal{E}(\alpha) \rangle  &=&  \frac{3}{8} \,
{\cos (\alpha )}^4 - \frac{2}{\pi} \, {\cos (\alpha )}^2\sin
(\alpha )\cr
 &+& \frac{1}{2} \, {\sin (\alpha )}^2.
\end{eqnarray}

\begin{figure}[h]
\begin{center}
\includegraphics[scale=0.3,angle=270]{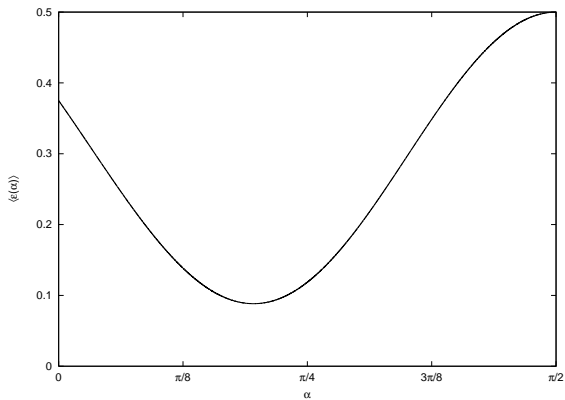}
\caption{Plot of $\langle \mathcal{E} (\alpha) \rangle$ as a
function of $\alpha$, $\alpha \in [0,\frac{\pi}{2}]$.}
\end{center}
\label{fig1}
\end{figure}

\noindent Figure 1 is a plot of the time averaged entanglement
$\langle \mathcal{E}(\alpha) \rangle$
as a function of the parameter $\alpha$. For
$|00\rangle \rightarrow |11\rangle$ ($\alpha=\frac{\pi}{2}$) we have
the optimal time evolution of two qubits from an initial symmetric
separable state into a final symmetric separable state orthogonal to
the initial one. In this case we have
$\langle \mathcal{E} \rangle = \frac{1}{2}$,
which is the maximum time-average entanglement in this family of
evolutions. On the other hand $|00\rangle \rightarrow
\frac{1}{\sqrt{2}}[|01\rangle+ |10\rangle]$ ($\alpha=0$) constitutes
an instance of the optimal evolution of a separable
state into a symmetric maximally entangled state. In
this case we have that $\langle \mathcal{E} \rangle = \frac{3}{8}$. Of particular
interest is the minimum of the time-average entanglement within the
family, which happens when $|00\rangle \rightarrow
\frac{1}{\sqrt{3}}[|01\rangle+ |10\rangle+|11\rangle]$
($\alpha=\arcsin\left(\frac{1}{\sqrt{3}}\right)$)
and gives $\langle \mathcal{E} \rangle = 0.088298$.\\


\noindent (ii). This constitutes a particular instance of the
optimal transformation of a symmetric maximally entangled state to
another symmetric maximally entangled state. The time dependent
entanglement and its average are, respectively,

\begin{equation}
\mathcal{E}(t) = \cos^2\left( \frac{\omega
t}{\hbar}\right), \,\,\,\,\,\,\,\,\,\,
\langle \mathcal{E} \rangle = \frac{1}{2}.
\end{equation}

\noindent In this particular case the same amount of time-averaged
entanglement is needed to transform a maximally entangled state into
another maximally entangled state as is needed to transform
a separable state into another separable state.\\

\noindent (iii) In this case the optimally evolving state is
maximally entangled at all times, consequently giving
$\mathcal{E}=1$.




\section{Typical Entanglement Properties of Time-Optimal
Evolutions of Two-Qubit Systems}

 \noindent Even when considering only two qubits and restricting
$|\psi_I\rangle$ and $|\psi_F\rangle$ to be symmetric and
orthogonal, the expression for $\langle \mathcal{E} \rangle$ becomes quite
involved, thus making an analytic treatment of certain properties of
the optimal evolution unpractical.

 In order to explore the typical entanglement
 features of optimal evolutions in two-qubits
 systems we are going to generate random pairs
 of symmetric (orthogonal) initial and final states,

 \ben \label{simestates}
 |\psi_I \rangle \, &=& \, c_1|00\rangle +
 c_2 \frac{1}{\sqrt{2}}\Bigl(|01\rangle + |10\rangle
 \Bigr)
 + c_3 |11\rangle, \cr
 |\psi_F \rangle \, &=& \, d_1|00\rangle +
 d_2 \frac{1}{\sqrt{2}}\Bigl(|01\rangle + |10\rangle
 \Bigr)
 + d_3 |11\rangle,
 \een

 \noindent
 with $\sum_i |c_i^2| = \sum_i |d_i^2| = 1$
 and $\sum_i c_i d_i^*=0$. We treat the case of
 symmetrical states separately because, as we
 shall presently see, in this case the necessity
 of entanglement to implement time-optimal dynamics
 is somewhat stronger than in the general case.
 However, an analysis similar to the one performed
 here for symmetrical states can be implemented
 for other particular families of states.
 We are going to consider the statistical
 distribution, and the mean and minimum values
 of the time averaged entanglement $\langle \mathcal{E}\rangle$
 associated with the time-optimal evolutions
 connecting the pairs of states $|\psi_I \rangle$
 and  $|\psi_F \rangle$. Interpreting the triplets
 $(c_1,c_2,c_3)$ and  $(d_1,d_2,d_3)$ as vectors
 in a three dimensional Hilbert space, we are going
 to generate the random states (\ref{simestates})
 by applying random $3\times 3$ unitary matrices
 $U$ to the vectors $(1,0,0)$ and $(0,1,0)$. The
 random matrices $U$ are generated uniformly according
 to the Haar measure (see \cite{BCPP02} and references
 therein). The corresponding probability distribution
 for the time averaged entanglement $\langle \mathcal{E}\rangle$
 is depicted in Figure 2 (solid line). This distribution,
 describing the (fractional) volume of state-space associated
 with different values of $\langle \mathcal{E} \rangle$,  has
 a lower cut-off at $\langle \mathcal{E} \rangle_{min} =
 0.03415330$, meaning that, in order to implement a time-optimal
 evolution between symmetric two-qubit states a finite,
 minimum amount of entanglement is needed. Such an evolution
 cannot be implemented without entanglement. On the other
 hand, there are time-optimal evolutions between
 symmetric states exhibiting a maximum time-averaged entanglement
 $\langle \mathcal{E} \rangle_{max}=1.0$. The mean
 value (over all possible optimal evolutions connecting
 symmetric states of two qubits is
 $\langle \mathcal{E} \rangle_{mean} = 0.5$.

\begin{figure}[h]
\begin{center}
\includegraphics[scale=0.3,angle=270]{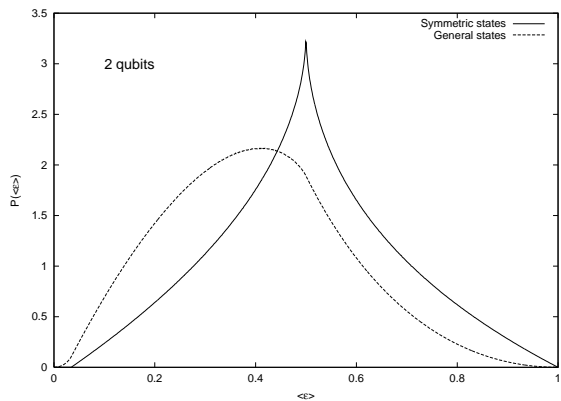}
\caption{Plot of the probability density function of the
time averaged entanglement $\langle \mathcal{E} \rangle$
associated with symmetric and general two-qubit states.}
\end{center}
\label{fig2}
\end{figure}

  We have also performed a similar calculation for time-optimal
 evolutions connecting two general (that is, not necessarily
 symmetrical) normalized and orthogonal states of two qubits,
 $|\psi_I \rangle  = a_1|00\rangle +
 a_2|01\rangle + a_3|10\rangle + a_4|11\rangle$
 and  $|\psi_F \rangle = b_1|00\rangle +
 b_2|01\rangle + b_3|10\rangle + b_4|11\rangle$.
 The vectors $(a_1,a_2,a_3,a_4)$ and $(b_1,b_2,b_3,b_4)$
 were obtained by acting with random unitary matrices $U$
 (generated uniformily according to the Haar measure) upon
 the vectors $(1,0,0,0)$ and $(0,1,0,0)$, respectively.
 The corresponding probability
 distribution for the $\langle \mathcal{E} \rangle$-values is
 shown in Figure 2 (dashed line). In this case there are
 optimal evolutions with arbitrarily small values of the time
 averaged entanglement. However, the probability density
 $P(\langle \mathcal{E} \rangle)$ approaches zero as
 $\langle \mathcal{E} \rangle \rightarrow 0$. Indeed, as we
 have already shown, the only optimal evolutions with
 $\langle \mathcal{E} \rangle=0$ are those of the form
 (\ref{defacto}) where one of the subsystems does not evolve.
 It can be clearly appreciated in Figure 2 that, as a general trend,
 more entanglement is involved in optimal evolutions between
 symmetric states than the one involved in general optimal
 evolutions (see also Table \ref{latablita}).

\section{Systems of Higher Dimensionality}

 We have also considered the connection between
entanglement and brachistochrone quantum evolution
for two-qutrit and three-qubit systems (as entanglement
measure for three-qubit states we used the average
bi-partite entanglement associated with the three
bi-partitions of the system into a qubit and a
two-qubit subsystem). The corresponding probability
densities for the time averaged entanglement
associated with brachistochrone evolutions connecting random
orthogonal states are depicted in Figure 3. It transpires
from this Figure that, both for two-qutrit and three-qubit
systems, a considerable amount of entanglement is involved
in typical brachistochrone evolutions.

\begin{figure}[h]
\begin{center}
\includegraphics[scale=0.5,angle=0]{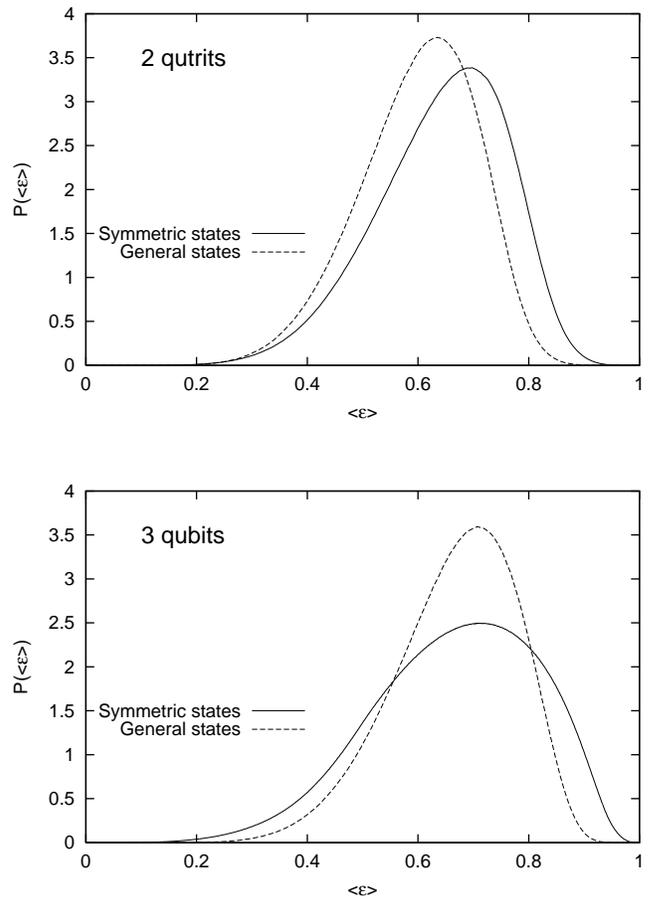}
\caption{Plot of the probability density function of the
time averaged entanglement associated with time optimal
quantum evolutions between symmetric and general orthogonal
states of two-qutrit and three-qubit states.}
\end{center}
\label{fig2}
\end{figure}

The minimum, maximum, mean and likeliest values of
the time averaged entanglement associated with brachistochrone
evolutions between orthogonal symmetric (S) and general (G) states (for
two-qubit, three-qubit and two-qutrit systems) are given in Table
\ref{latablita}.

\vskip 0.2cm
\begin{table}[h]
\begin{center}
\begin{tabular}{|c|c|c|c|c|c|c|}
\hline
&\multicolumn{2}{c|}{2 qubits}&\multicolumn{2}{c|}{3 qubits}&\multicolumn{2}{c|}{2 qutrits}\\
\hline
& \ S \  & \ G \ & \ S \ & \ G \ & \ S \ & \ G \ \\
\hline
$E_{min}$ & 0.0342 & 0.0000    & 0.0304 & 0.0000 & 0.0256 & 0.0000 \\
\hline
$E_{max}$ & 1.0000 & 1.0000 & 1.0000 & 1.0000 & 0.9914 & 0.9915 \\
\hline
$E_{mean}$ & 0.5000   & 0.4000 & 0.6667 & 0.6667 & 0.6429 & 0.5999 \\
\hline
$E_{likeliest}$ & 0.50 & 0.41 & 0.71 & 0.71 & 0.69 & 0.64 \\
\hline
\end{tabular}
\end{center}
\caption{Minimum, maximum, mean and likeliest
time-averaged entanglement associated with
brachistochrone evolutions of systems of
2 qubits, 3 qubits, and 2 qutrits.}
\label{latablita}
\end{table}

\vskip -1cm

\section{Conclusions}

We have investigated the role of entanglement in
time-optimal (brachistochrone) evolution of
composite quantum systems. We proved that,
except for trivial cases in which only one of the
subsystems is actually evolving, {\it brachistochrone
quantum evolution between orthogonal states cannot
be implemented without entanglement}. We
studied in detail, for two-qubits, three-qubits
and two-qutrits systems, the amount of entanglement
(as measured by the time averaged entanglement)
involved in typical brachistochrone evolutions
connecting orthogonal initial and final states.
In all cases we found that a considerable amount
of entanglement is needed in order to implement
brachistochrone evolutions. The present approach
to the study of the connection between entanglement
and quantum time-optimal evolution may be regarded
as ``global" in the sense that (for a given system)
it is not based upon the {\it separate} analysis of
the the dynamics generated by different
possible Hamiltonians. If we consider, for instance,
a two-qubits system, the global approach
shows that, in general, entanglement is a
necessary resource to implement  optimal quantum
evolutions. Previous approaches, on the contrary,
would require a separate treatment of each different
Hamiltonian of the system at hand (say, a
two-qubit system).

The two alluded  strategies, the ``global" and
the ``Hamiltonian-specific" one, actually
complement each other. The global approach,
based upon the brachistochrone evolution, establishes
in a general and unified way that there is a connection between
entanglement and optimal quantum evolution between
orthogonal pure states: {\it considering
at the same time (for a given system) all possible optimum
quantum evolutions it is seen that most of them involve
a considerable amount of entanglement}. On the
other hand, the Hamiltonian-specific treatment
permits, for each specific quantum Hamiltonian,
a more detailed analysis of the aforementioned
connection \cite{GLM03a,GLM03b,BCPP05,BCPP06,ZPPC07}.

Summing up, in the present work we have established a
definite connection between entanglement and brachistochrone
evolutions connecting pairs of pure orthogonal states.
Arguably, time-optimal evolutions between {\it pure and
orthogonal states} are (at least from the conceptual
point of view) the most important ones, since the initial
and the final state are fully distinguishable and the
alluded evolution can be identified with one
elementary information-processing step \cite{ML98}.
However, it is possible to incorporate within the present
approach brachistochrone evolutions connecting non-orthogonal
pure states, since these evolutions are perfectly well
defined \cite{BH06}. Moreover, it would also be possible to explore
the case of optimal time-evolution between mixed states,
using the techniques advanced in \cite{CHKO07}. We plan
to investigate these extensions of the ideas
advanced here in a future contribution.

\vskip 0.5cm

\noindent
{\bf Acknowledgements.}
We would like to thank the anonimous referees for their
useful suggestions that helped to clarify and improve
the present contribution. The financial
assistance of the National Research Foundation (NRF; South African
Agency) towards this research is hereby acknowledged. Opinions
expressed and conclusions arrived at, are those of the authors and
are not necessarily to be attributed to the NRF. This work was
partially supported by the MEC grant FIS2005-02796 (Spain) and FEDER
(EU), by the Government of Balearic Islands, and by CONICET
(Argentine Agency). A. Borr\'as acknowledges support from FPU grant
AP-2004-2962 (MEC-Spain).


\end{document}